\documentclass{ws-procs961x669}            % default, citations in superscript

\usepackage{tensor}
\newcommand{\leri}[1]{\left(#1\right)}

\begin{document}
\title{Big-Bounce in projectively invariant Nieh-Yan models: the Bianchi I case}

\author{Flavio Bombacigno}

\address{Departament de F\'{i}sica Teòrica and IFIC, Centro Mixto Universitat de València - CSIC, Universitat de València, Burjassot 46100, València, Spain\\
E-mail: flavio.bombacigno@ext.uv.es}

\author{Simon Boudet}

\address{Dipartimento di Fisica, Universit\`{a} di Trento,\\Via Sommarive 14, I-38123 Povo (TN), Italy\\
and\\
Trento Institute for Fundamental Physics and Applications (TIFPA)-INFN,\\Via Sommarive 14, I-38123 Povo (TN), Italy\\
E-mail: simon.boudet@unitn.it}

\author{Gonzalo J. Olmo}
\address{Departament de F\'{i}sica Teòrica and IFIC, Centro Mixto Universitat de València - CSIC, Universitat de València, Burjassot 46100, València, Spain\\
E-mail:gonzalo.olmo@uv.es}

\author{Giovanni Montani}
\address{Physics Department, ``Sapienza'' University of Rome,\\ P.le Aldo Moro 5, 00185 (Roma), Italy\\
and\\
ENEA, Fusion and Nuclear Safety Department,\\ C. R. Frascati,
 	Via E. Fermi 45, 00044 Frascati (Roma), Italy\\
 E-mail: giovanni.montani@enea.it}

\begin{abstract}
We show that the Nieh-Yan topological invariant breaks projective symmetry and loses its topological character in presence of non vanishing nonmetricity. 
The notion of the Nieh-Yan topological invariant is then extended to the generic metric-affine case, defining a generalized Nieh-Yan term, which allows to recover topologicity and projective invariance, independently.
As a concrete example a class of modified theories of gravity is considered and its dynamical properties are investigated in a cosmological setting. In particular, bouncing cosmological solutions in Bianchi I models are derived. Finite time singularities affecting these solutions are analysed, showing that the geodesic completeness and the regular behavior of scalar perturbations in these space-times are not spoiled.
\end{abstract}


\bodymatter

\section{Introduction}
The theory of General Relativity (GR) \cite{WaldR.M.1984,MisnerC.W.2017} relies on the geometric interpretation of the gravitational field, described in terms of a metric tensor and a connection on a pseudo-Riemannian manifold. Both GR and many alternative theories of gravity are based on a metric formulation, in which the connection is given by the symmetric and metric compatible Levi-Civita connection, which is completely determined by the metric and its derivatives. An alternative formulation for geometric theories of gravity consists in adopting the metric-affine paradigm, in which the metric tensor and the connection are considered as independent variables. In this approach, symmetry and metric compatibility of the connection are not imposed a priori, resulting in the presence of torsion and nonmetricity, respectively. Well known examples of metric-affine theories are Ricci based gravity \cite{Afonso:2018bpv,Afonso:2018hyj}, Palatini $f(R)$ theory \cite{Olmo2011}, quadratic gravity \cite{Lobo:2013adx}, Born-Infeld-type models \cite{BeltranJimenez:2017doy}, general teleparallel models \cite{BELTRANJIMENEZ2020135422}, generalized hybrid metric-Palatini gravity \cite{Rosa:2021lhc,Harko:2011nh,Capozziello:2015lza,Bombacigno:2019did,Bronnikov:2019ugl} and metric-affine extension of higher order theories \cite{Borunda:2008kf,Janssen:2019doc,Janssen:2019uao,Percacci:2019hxn}.\\
The metric-affine approach plays a crucial role also in one of the current attempts to quantize gravity, i.e. loop quantum gravity (LQG) \cite{Rovelli2004,Thiemann2007}, where GR is reformulated in terms of a gauge $SU(2)$ connection (Ashtekar-Barbero-Immirzi connection) and its conjugate momentum, the densitized triad \cite{Ashtekar1986,Ashtekar1987,Ashtekar1989,Ashtekar1992}. This formulation, indeed, can be derived \cite{Date2009} by including an additional contribution to the first order (Palatini) action of GR, namely the Nieh-Yan (NY) topological invariant \cite{Nieh1982,Nieh2007} (the Holst term \cite{Holst1996} can be used as well). The NY term was discovered in the context of Riemann-Cartan theory (where nonmetricity is set to zero) and its main property is topologicity: it reduces to a boundary term without affecting the field equations at all. This additional term is driven by the so called Immirzi parameter $\beta$ \cite{Immirzi:1996di,Immirzi1997}, which concurs in the definition of the Ashtekar variables and is related to a quantization ambiguity \cite{Rovelli:1997na}. Attempts to address this issue led to the proposal of considering the Immirzi parameter as a new fundamental field \cite{Taveras2008,Calcagni2009,Bombacigno2016}, an idea that has been later developed within several different contexts \cite{Veraguth2017,Wang2018,Bombacigno2019,Bombacigno2018,Wang2020,Iosifidis:2020dck,BOMBACIGNO2021115281,Langvik:2020nrs,Taveras2008,Bombacigno2016,BombacignoFlavioandMontani2019,PhysRevD.103.084034,PhysRevD.81.125015,PhysRevD.91.085017,Mercuri2009a,Bombacigno2016,BOMBACIGNO2021115281}. The promotion of such constant parameter to a dynamical field is usually pursued ``by hand'', substituting $\beta\rightarrow \beta(x)$ in the Lagrangian and possibly adding a potential term $V(\beta)$. 
\\More recently, beside LQG the NY term has been studied in the context of teleparallel gravity \cite{Li_2020} and in condensed matter physics \cite{PhysRevResearch.2.033269,Nissinen2019,PhysRevB.101.125201,liu2021phonon}.\\
Another important property we will focus on, is projective invariance \cite{Afonso:2017bxr,Iosifidis:2019fsh}, which has recently been shown to be of crucial importance in metric-affine theories since the breaking of this symmetry can give rise to dynamical instabilities \cite{BeltranJimenez2019}. In this regard, we want to stress that the NY term breaks this symmetry. This feature has always been neglected in literature and a revision of previous formulations seems necessary. Moreover, as will be shown in the following, the topological character of the NY term is also lost when nonmetricity is included.\\
The approach followed in this note is grounded on the choice of recovering these features from the very beginning in the action, without imposing any restriction on the affine connection. After a formal discussion, we will implement the gravitational model in a cosmological setting. In particular, we investigate Bianchi I models  \cite{Montani:2011zz,Corichi:2009pp,Kamenshchik:2017ojc}, focusing our attention on the emergence of a classical bouncing cosmology \cite{Montani:2021yom,Montani:2018bxv,Cianfrani:2012gv,Moriconi:2017bvs,Giovannetti:2021bqh,BombacignoFlavioandMontani2019,Barragan:2010qb,Barragan:2009sq}.

\section{The role of nonmetricity in the Nieh-Yan term}\label{section 3}
In Einstein-Cartan theory the NY term \cite{Nieh1982,Nieh2007} is explicitly defined as
\begin{align}
    NY &\equiv\frac{1}{2}\varepsilon^{\mu\nu\rho\sigma}\leri{\frac{1}{2}T\indices{^\lambda_{\mu\nu}}T_{\lambda\rho\sigma}-R_{\mu\nu\rho\sigma}} \ ,
    \label{NY0}
\end{align}
where
\begin{equation}
    R\indices{^\rho_{\mu\sigma\nu}}=\partial_\sigma\Gamma\indices{^\rho_{\mu\nu}}-\partial_\nu\Gamma\indices{^\rho_{\mu\sigma}}+\Gamma\indices{^\rho_{\tau\sigma}}\Gamma\indices{^\tau_{\mu\nu}}-\Gamma\indices{^\rho_{\tau\nu}}\Gamma\indices{^\tau_{\mu\sigma}},
\end{equation}
is the Riemann tensor built with the independent connection and
\begin{equation}
    T\indices{^\mu_{\nu\rho}}= \Gamma\indices{^\mu_{\nu\rho}}- \Gamma\indices{^\mu_{\rho\nu}}
\end{equation}
is the torsion tensor.
\noindent The starting point of our discussion is the observation that for a non-vanishing nonmetricity tensor $Q_{\mu\nu\rho}=-\nabla_\mu g_{\nu\rho}$, the NY term \eqref{NY0} is spoilt of its topological character. Indeed, extracting the nonriemmanian part of the Riemann tensor leads to\cite{PhysRevD.103.124031}
\begin{align}
    NY &=-\frac{1}{2}\bar{\nabla}\cdot S-\frac{1}{2}\varepsilon^{\mu\nu\rho\sigma}T\indices{^\lambda_{\mu\nu}}Q\indices{_{\rho\sigma\lambda}},
    \label{NY}
\end{align}
where $\bar{\nabla}_\mu$ is built with the Levi-Civita connection $\bar{\Gamma}\indices{^\mu_{\nu\rho}}$ and
\begin{equation}
S_{\mu} \equiv \varepsilon_{\mu\nu\rho\sigma}T^{\nu\rho\sigma}.
\end{equation}
Therefore, when $Q_{\rho\mu\nu}\neq 0$, the Nieh-Yan term does not simply reduce to the divergence of a vector, and the appearance of nonmetricity spoils the topologicity. Let us now consider the behavior of \eqref{NY0} under projective transformations of the connection, namely
\begin{equation}
\tilde{\Gamma}\indices{^\rho_{\mu\nu}}=\Gamma\indices{^\rho_{\mu\nu}}+\delta\indices{^\rho_\mu}\xi_\nu,
    \label{projective}\end{equation}
It can easily be shown that \eqref{NY0} is also not invariant under projective transformations, since
\begin{equation}
    \frac{1}{4}\varepsilon^{\mu\nu\rho\sigma}\tilde{T}\indices{^\lambda_{\mu\nu}}\tilde{T}_{\lambda\rho\sigma}-\frac{1}{4}\varepsilon^{\mu\nu\rho\sigma}T\indices{^\lambda_{\mu\nu}}T_{\lambda\rho\sigma}=-S_\mu\xi^\mu.
    \label{projective rule TT}
\end{equation}
Now, by looking at \eqref{NY}, we point out that a newly topological Nieh-Yan term can be recovered by simply setting
\begin{equation}
    NY^{*}\equiv NY+\frac{1}{2}\varepsilon^{\mu\nu\rho\sigma}T\indices{^\lambda_{\mu\nu}}Q\indices{_{\rho\sigma\lambda}}.
\end{equation}\label{NY top}
We note that projective invariance is now enclosed as well, since
\begin{equation}
        \frac{1}{2}\varepsilon^{\mu\nu\rho\sigma}\tilde{T}\indices{^\lambda_{\mu\nu}}\tilde{Q}\indices{_{\rho\sigma\lambda}}-\frac{1}{2}\varepsilon^{\mu\nu\rho\sigma}T\indices{^\lambda_{\mu\nu}}Q\indices{_{\rho\sigma\lambda}}=+S_\mu\xi^\mu,
\end{equation}
which exactly cancels out \eqref{projective rule TT}. We stress, however, that projective invariance is not strictly related to topologicity, and suitable generalizations of \eqref{NY top} breaking up only with the latter can be actually formulated. Let us consider, for instance, the following modified Nieh-Yan term
\begin{equation}
    NY_{gen}\equiv\frac{1}{2}\varepsilon^{\mu\nu\rho\sigma}\leri{\frac{\lambda_1}{2}T\indices{^\lambda_{\mu\nu}}T_{\lambda\rho\sigma}+\lambda_2\, T\indices{^\lambda_{\mu\nu}}Q\indices{_{\rho\sigma\lambda}}-R_{\mu\nu\rho\sigma}},
    \label{NY general}
\end{equation}
where we introduced the real parameters $\lambda_1,\,\lambda_2$. In this case the term \eqref{NY general} transforms under a projective transformation as
\begin{equation}
    NY_{gen}\rightarrow NY_{gen}-(\lambda_1-\lambda_2)\xi^\mu S_\mu,
\end{equation}
so that by setting $\lambda_1=\lambda_2$ we can recover again projective invariance, despite topologicity being in general violated if $\lambda_1=\lambda_2\neq 1$, since
\begin{equation}
     NY_{gen}=-\frac{1}{2}\bar{\nabla}\cdot S+\frac{ (\lambda_1-1)}{4}\varepsilon^{\mu\nu\rho\sigma}T\indices{^\lambda_{\mu\nu}}T_{\lambda\rho\sigma}+\frac{ (\lambda_2-1)}{2}\varepsilon^{\mu\nu\rho\sigma} T\indices{^\lambda_{\mu\nu}}Q\indices{_{\rho\sigma\lambda}}.
\end{equation}
In the following, we will consider the general form \eqref{NY general}, which by a suitable choice of the parameters $\lambda_{1,2}$ can reproduce all  known actions usually studied in literature, as the Holst ($\lambda_1=\lambda_2=0$) or the standard Nieh-Yan \eqref{NY} ($\lambda_1=1,\,\lambda_2=0$) terms.

\section{Generalized Nieh-Yan models}\label{section 4}
As a specific gravitational model featuring the generalized NY term we consider an action defined by a general function of two arguments, the Ricci scalar and the generalized NY term \eqref{NY general}:
\begin{equation}
    S_g=\frac{1}{2\kappa}\int d^4x \sqrt{-g}\,F(R,NY_{gen}),
    \label{action general ny}
\end{equation}
Now, performing the transformation to the Jordan frame leads to the scalar tensor representation
\begin{equation}
    S_g=\frac{1}{2\kappa}\int d^4x \sqrt{-g}\leri{\phi R+\beta NY_{gen}-W(\phi,\beta)},
    \label{action general scalar tensor ny}
\end{equation}
with $\phi\equiv\frac{\partial F}{\partial R},\,\beta\equiv\frac{\partial F}{\partial NY_{gen}}$ and $W\equiv\phi R(\phi,\beta)+\beta NY_{gen}(\phi,\beta)-F(\phi,\beta)$. 
\\ \noindent The scalar field $\beta$ can be identified with the Immirzi field, which acquires in this way a dynamical character without the need of introducing this feature by hand in the action. Moreover, this formulation offers a viable mechanism to produce an interaction term $W(\phi,\beta)$ as well.
Now, the field equation for the connection are obtained varying \eqref{action general scalar tensor ny} with respect to $\Gamma\indices{^\mu_{\nu\rho}}$. For the full set of equations the reader may cosnult \cite{PhysRevD.103.124031}, while here we are interest in the following contraction
\begin{equation}
  \delta^\lambda_\nu  \frac{\delta S_g}{\delta \Gamma\indices{^\lambda_{\nu\mu}}}=0
\end{equation}
which leads to
\begin{equation}
    (\lambda_1 - \lambda_2)S^\mu =0.
\end{equation}
This implies that the features of the solutions depend on the parameters $\lambda_1$ and $\lambda_2$, and when projective invariance is broken ($\lambda_1\neq\lambda_2$) one is compelled to set $S_\mu=0$.
In this case, \eqref{NY general} can be re-expressed as
\begin{equation}\label{NYgen components}
    NY_{gen}=-\frac{1}{2}\bar{\nabla}\cdot S-\frac{(1-\lambda_1)}{3}S^\mu T\indices{^\rho_{\mu\rho}}-\frac{(1-\lambda_2)}{2}S^\mu Q\indices{^\rho_{\mu\rho}},
\end{equation}
implying that the generalized Nieh-Yan term \eqref{NY general} is identically vanishing on half-shell. In other words, terms violating projective invariance are harmless along the dynamics. This can be further appreciated deriving the effective scalar tensor action stemming from \eqref{action general scalar tensor ny}, when the solutions of the full set of connection field equations are plugged in it. Explicit calculations (see \cite{PhysRevD.103.124031} for details)
lead to
\begin{equation}
    S=\frac{1}{2\kappa}\int d^4x \sqrt{-g}\leri{\phi \bar{R}+\frac{3}{2\phi}\bar{\nabla}_\mu\phi\bar{\nabla}^\mu\phi-W(\phi,\beta)},
    \label{NY scalar tensor}
\end{equation}
where $\bar{R}$ is the Ricci scalar of the Levi-CIvita connection. This resembles the form of a Palatini $f(R)$ theory, with a potential depending on the Immirzi field as well. The equation for the latter, i.e.
\begin{equation}
    \frac{\partial W(\phi,\beta)}{\partial \beta}=0,
    \label{NY beta}
\end{equation}
fixes its form in terms of the remaining scalar field: $\beta=\beta(\phi)$. Then, using the trace of the metric field equations, variation of \eqref{NY scalar tensor} with respect to $\phi$ results in the usual structural equation featuring Palatini $f(R)$ theories \cite{Olmo2011}, i.e.
\begin{equation}\label{structural NY}
\left[2W(\phi,\beta)-\phi\frac{\partial W(\phi,\beta)}{\partial\phi}\right]_{\beta=\beta(\phi)}=\kappa T,
\end{equation}
where $T$ is the trace of the stress energy tensor of matter. This implies that the dynamics of the scalaron $\phi$ is frozen as well, and completely determined by $T$. Conditions \eqref{NY beta} and \eqref{structural NY} then guarantee that the scalar fields $\phi,\;\beta$ are not propagating degrees of freedom, and reduce to constants in vacuum, where the theory is stable and the breaking of projective invariance does not lead to ghost instabilities, in contrast to \cite{BeltranJimenez2019}.\\
If $\lambda_1=\lambda_2\equiv\lambda$, instead, the projective invariance of the model can be used to get rid of one vector degree of freedom, which can be set to zero properly choosing the vector $\xi_\mu$. A convenient choice consists in setting $\xi_\mu=-\frac{1}{2}Q\indices{^\rho_{\mu\rho}}$, which allows to deal only with torsion in the connection field equations. The effective action stemming from \eqref{action general scalar tensor ny} then reads (see \cite{PhysRevD.103.124031})
\begin{equation}
    S=\frac{1}{2\kappa}\int d^4 x \sqrt{-g} \leri{\phi \bar{R}+\frac{3}{2\phi}\bar{\nabla}_\mu\phi\bar{\nabla}^\mu\phi-\frac{3\phi}{2}\frac{1}{\phi^{2\lambda}+(1-\lambda)^2\psi^2}\bar{\nabla}_\mu\psi\bar{\nabla}^\mu\psi-V(\phi,\psi)},
    \label{effective action projective preserving}
\end{equation}
where we used the transformation $\psi\equiv\beta\phi^{\lambda-1}$ and redefined the potential as $V(\phi,\psi)=W(\phi,\psi \phi^{1-\lambda})$. In general, the Immirzi field is expected to be a well-behaved dynamical degree of freedom, since in the Einstein frame action, defined by the conformal rescaling $\tilde{g}_{\mu\nu}=\phi\,g_{\mu\nu}$, the kinetic term for the Immirzi field takes the form
\begin{equation}
    -\frac{3}{2}\frac{\tilde{g}^{\mu\nu}\nabla_\mu\psi\nabla_\nu\psi}{\phi^{2\lambda}+(1-\lambda)^2\psi^2}.
    \label{noghost}
\end{equation}
Since $\phi^{2\lambda}+(1-\lambda)^2\psi^2>0$ for every value of $\phi,\,\psi$ and $\lambda$, \eqref{noghost} has always the correct sign and no ghost instability arise \cite{Olmo:2005hc}.\\
Let us end this section with a remark on how previous results with vanishing nonmetricity can properly be recovered. In particular, the Riemann-Cartan structure of \cite{Calcagni2009,Mercuri2006,Mercuri2008,Mercuri2009,Mercuri2009a,Bombacigno2016,Bombacigno2018,Bombacigno2019} can be replicated by inserting in \eqref{action general ny} the condition of vanishing nonmetricity with a Lagrange multiplier, i.e. adding to the Lagrangian a term $l^{\rho\mu\nu}Q_{\rho\mu\nu}$, with $l^{\rho\mu\nu}=l^{\rho\nu\mu}$. Then, results of \cite{Calcagni2009,Mercuri2006,Mercuri2008,Mercuri2009,Mercuri2009a,Bombacigno2016,Bombacigno2018,Bombacigno2019} are simply obtained by setting $\lambda_1=1$. The fact that the usual Einstein-Cartan NY invariant and related models are recovered in this way, supports the correctness of our generalization of the NY term, with respect to other possible generalizations.

\section{Big bounce in Bianchi I cosmology}\label{section 6}
In this section we consider dynamical models, i.e. those described by \eqref{effective action projective preserving}, which are characterized by a dynamical Immirzi field and look for cosmological solutions in Bianchi I spacetimes. In particular, we will be interested in obtaining solutions characterized by a bouncing behavior for the universe volume, thanks to which the big bang singularity is regularized in favour of a big bounce scenario.

Let us start from the equations of motion for the metric and scalar fields which are given by
\begin{align}
    \bar{G}_{\mu\nu}&=\frac{\kappa}{\phi}T_{\mu\nu}+\frac{1}{\phi}\leri{\bar{\nabla}_\mu\bar{\nabla}_\nu-g_{\mu\nu}\bar{\Box}}\phi-\frac{3}{2\phi^2}\bar{\nabla}_\mu\phi\bar{\nabla}_\nu\phi+\frac{3}{2}\frac{\bar{\nabla}_\mu\psi\bar{\nabla}_\nu\psi}{\phi^{2\lambda}+(1-\lambda)^2\psi^2}\nonumber\\
    &+\frac{1}{2}g_{\mu\nu}\leri{\frac{3(\bar{\nabla}\phi)^2}{2\phi^2}-\frac{3}{2}\frac{(\bar{\nabla}\psi)^2}{\phi^{2\lambda}+(1-\lambda)^2\psi^2}-\frac{V(\phi,\psi)}{\phi}},\label{equation DI metric}
    \\
     &2V(\phi,\psi)-\phi\frac{\partial V(\phi,\psi)}{\partial\phi}+\frac{3\lambda \phi^{2\lambda+1}}{\leri{\phi^{2\lambda}+(1-\lambda)^2\psi^2}^2}(\bar{\nabla}\psi)^2=\kappa T,
    \label{structural DI}
    \\
    &\bar{\Box}\psi-\frac{(1-\lambda)^2\psi}{\phi^{2\lambda}+(1-\lambda)^2\psi^2}(\bar{\nabla}\psi)^2+\leri{1-\frac{2\lambda\phi^{2\lambda}}{\phi^{2\lambda}+(1-\lambda)^2\psi^2}}\bar{\nabla}_\mu\ln\phi\bar{\nabla}^\mu\psi=\frac{\partial V(\phi,\psi)}{3\partial\psi}.
\end{align}
As will be shown, cosmological solutions can be found for projective invariant models ($\lambda=1$), and restricting to potentials of the form $V(\phi,\psi)=V(\phi)$. To this end, we consider the metric for a Bianchi I flat spacetime, i.e.
\begin{equation}
    ds^2=-dt^2+a(t)^2dx^2+b(t)^2dy^2+c(t)^2dz^2,
    \label{bianchi metric}
\end{equation}
which represents a homogeneous but anisotropic spacetime, with three different scale factors $a(t),b(t),c(t)$. We include matter in the form of a perfect fluid with stress energy tensor
\begin{equation}
    T_{\mu\nu}=\text{diag}(\rho, a^2p,b^2p,c^2p),
    \label{perfect fluid tensor}
\end{equation}
where $\rho$ is the energy density and $p$ the pressure.
Assuming the equation of state $p=w\rho$, conservation of the stress energy tensor implies
\begin{equation}\label{energy density}
    \rho(t)=\frac{\mu^2}{(abc)^{w+1}},
\end{equation}
where $\mu^2$ is a constant. Now, considering a Starobinsky quadratic potential \cite{STAROBINSKY198099}
\begin{equation}
    V(\phi)=\frac{1}{\alpha}\leri{\frac{\phi-1}{2}}^2,
    \label{potential quadratic}
\end{equation}
it can be shown that the equations for the scalar fields yield\footnote{A dot denotes derivatives with respect to time.} 
\begin{equation}
    \dot{\psi}=\frac{k_0\phi}{v}, \qquad\qquad\qquad \phi=\frac{v^2 f(v)}{6\alpha k_0^2+v^2},
    \label{solution immirzi DI hy}
\end{equation}
in terms of the volume-like variable $v\equiv abc$ and the function $f(v)\equiv 1-2\alpha\kappa(3w-1)\rho(v)$, while $k_0$ is an integration constant. Regarding the metric field equations, lengthy computations\cite{PhysRevD.103.124031} allow to rewrite the tt component in the form of a modified Friedmann equation:
\begin{equation}
    H^2\equiv\leri{\frac{\dot{v}}{3v}}^2=\frac{\frac{\kappa}{3}\leri{\frac{\mu_I^2}{v^2}+\frac{\rho}{\phi }+\frac{\mu_{AN}^2}{\phi^2v^2}}+\frac{V(\phi)}{6\phi}}{\leri{1+\frac{3v}{2}\frac{d}{dv}\ln\phi}^2},
    \label{hubble function ny}
\end{equation}
where $\mu_A^2$ and $\mu^2_I$ are constants, representing the anisotropy density parameter   and the energy density parameter of the Immirzi field, respectively. We note that the r.h.s is a rational function of the volume alone.\\
In the following we will take into account 
dust and radiation as matter contributions, specified by the choices $w=0$, $\mu=\mu_D$ and $w=1/3$, $\mu=\mu_R$ in \eqref{energy density}, respectively. Then, bouncing solutions can be derived for $\alpha<0$ integrating Eq.\eqref{hubble function ny} for $v(t)$, which then yields the scalar fields behavior via \eqref{solution immirzi DI hy}.

\subsection{Vacuum case}
\begin{figure}[h]
\centering
\includegraphics[width=2.5in]{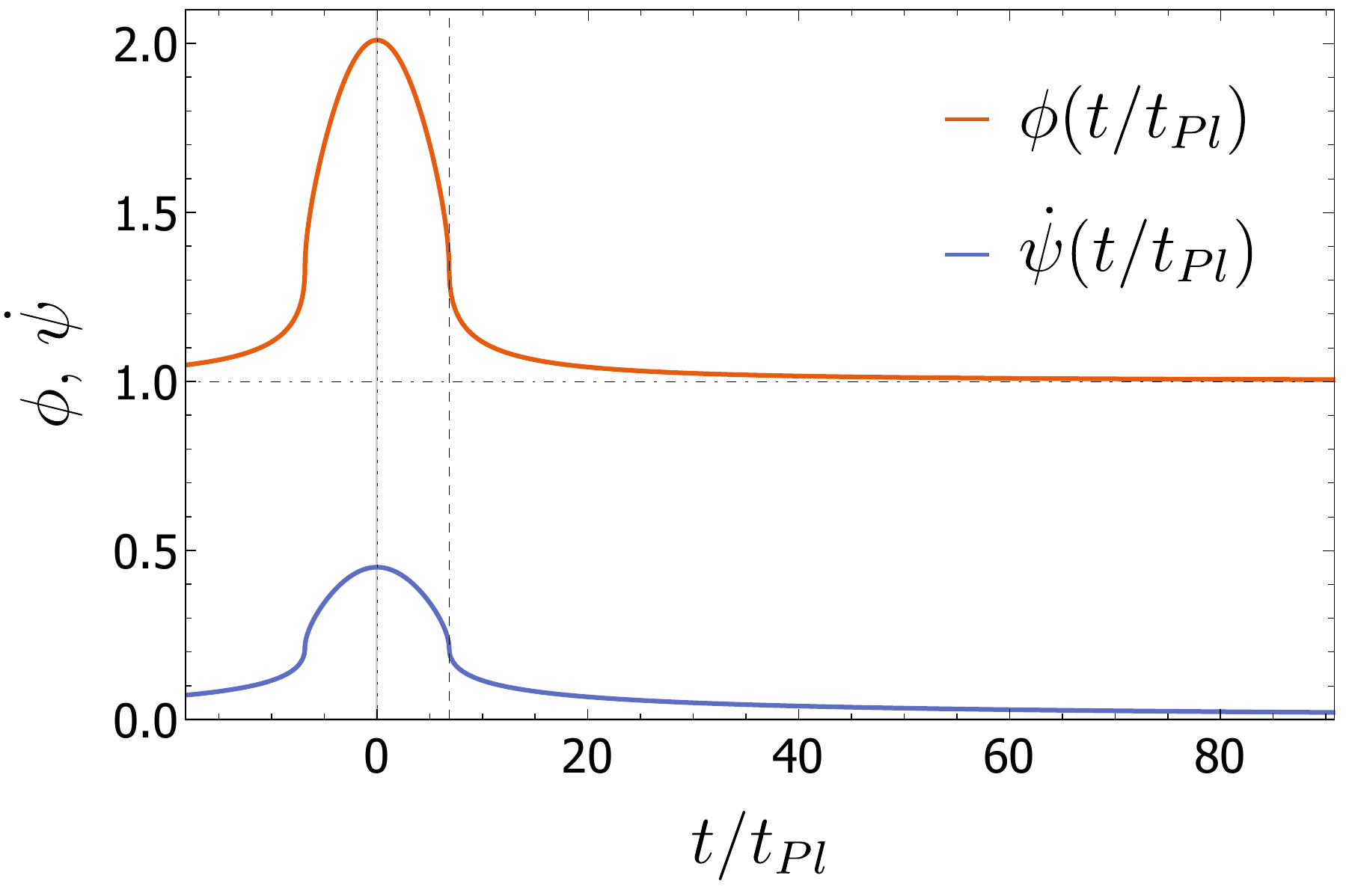}
\caption{Scalaron $\phi$ and Immirzi field derivative $\dot{\psi}$ for $\alpha=-5/3$, $\mu_I=\sqrt{3}$, $\mu_A=0.2 \mu_I$ as a function of $t/t_{Pl}$. }
\label{fig: PhiPsi}
\end{figure}

It is convenient to first focus on the vacuum case, where $f(v)=1$ and $\mu_D=\mu_R=0$. In this case we found the big bounce in the volume variable depicted in Fig.~\ref{fig: PlotV}, where also the behavior of each scale factor is shown.
\def\figsubcap#1{\par\noindent\centering\footnotesize(#1)}
\begin{figure}[h]%
\begin{center}
  \parbox{2.1in}{\includegraphics[width=2in]{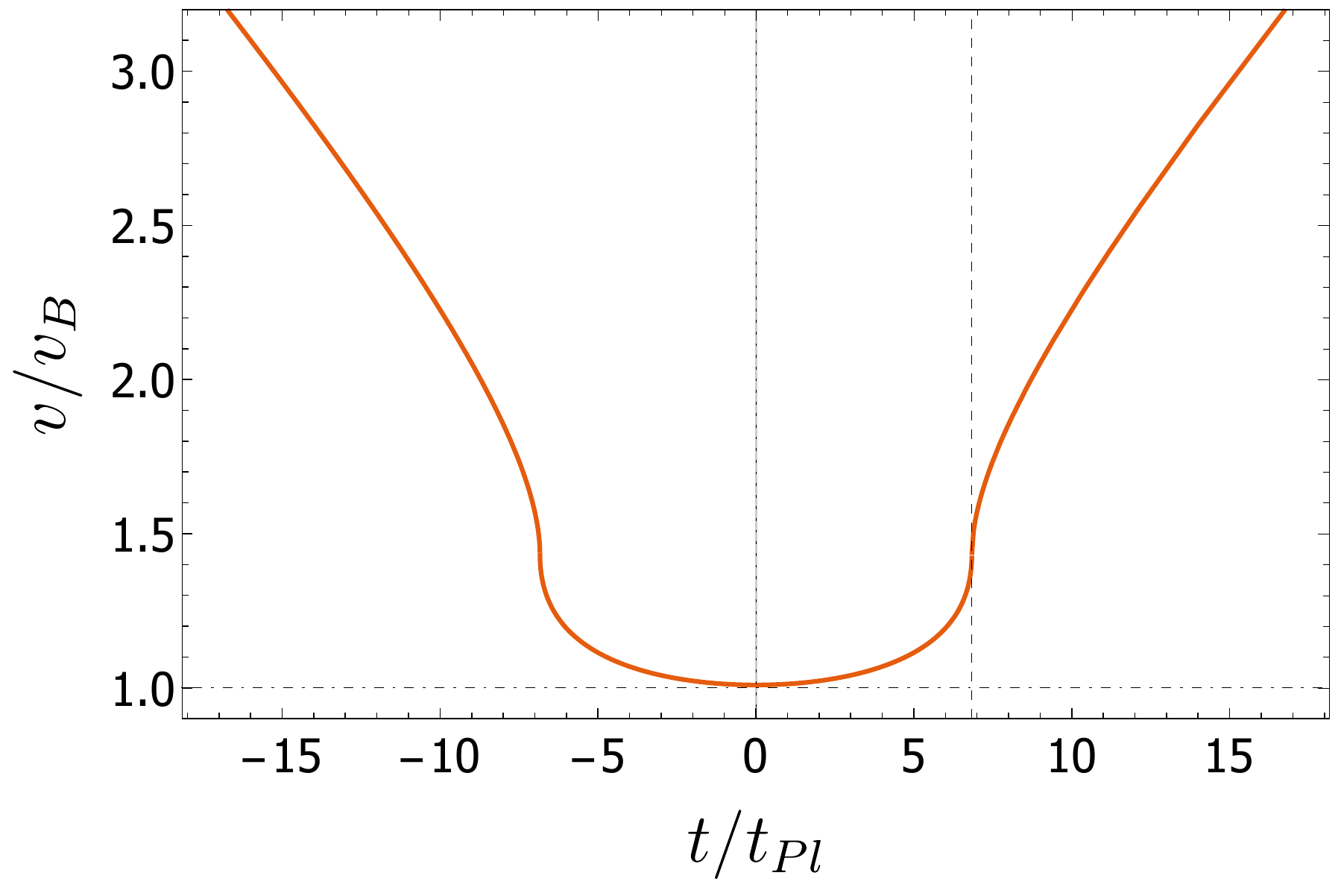}\figsubcap{a}}
  \hspace*{4pt}
  \parbox{2.1in}{\includegraphics[width=2in]{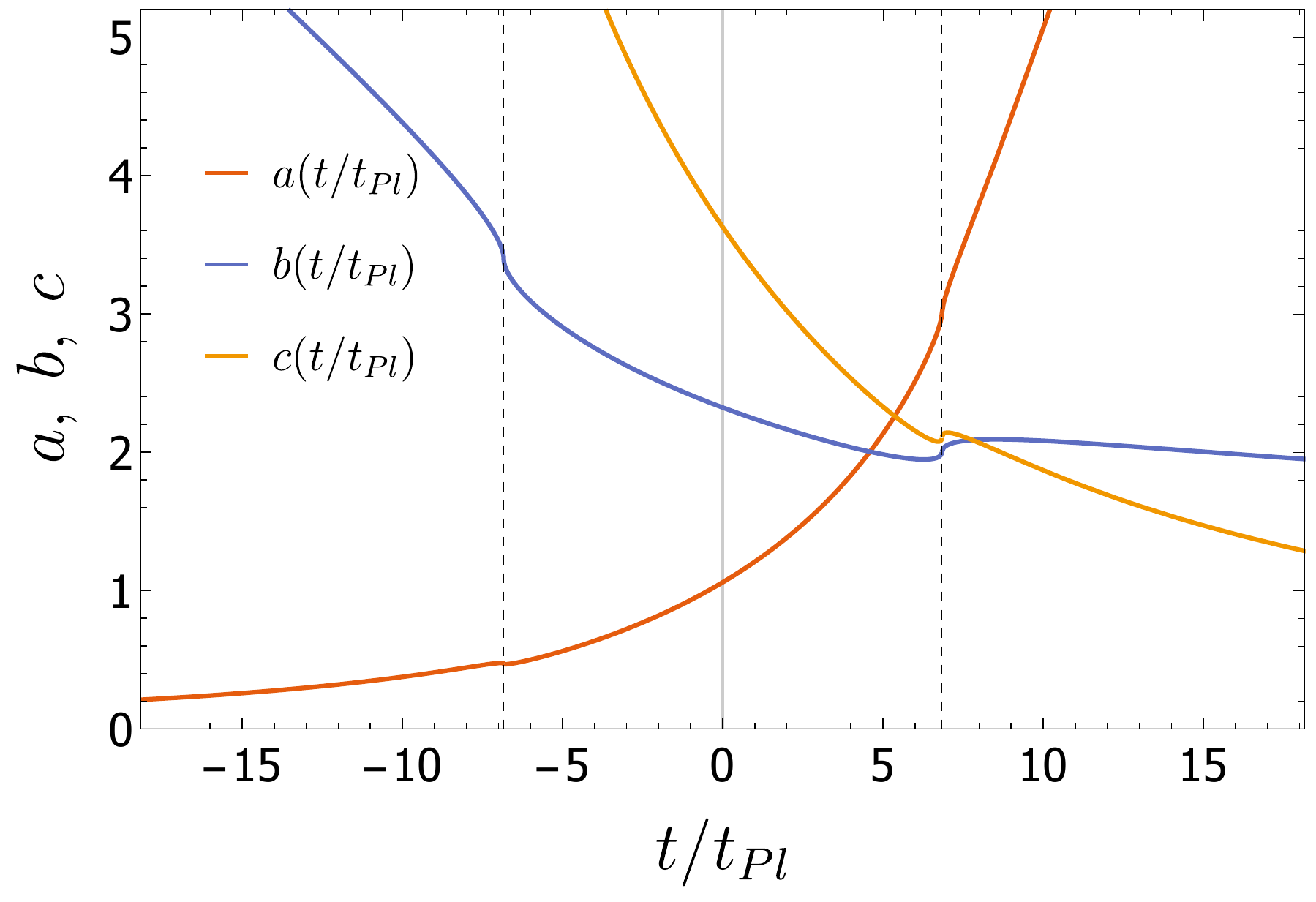}\figsubcap{b}}
  \caption{Numerical solutions for $\alpha=-5/3$, $\mu_I=\sqrt{3}$, $\mu_A=0.2 \mu_I$ as a function of $t/t_{Pl}$. Dotted and dashed lines represent where bounce and future time singularity happen, respectively. The bounce is centered at the origin of time for convenience, and the values of the parameters are chosen in order to yield graphs that display features in a clear fashion. (a) Universe volume normalised to $v_B$. (b) Scale factors.}%
  \label{fig: PlotV}
\end{center}
\end{figure}
We note that the volume is affected by a future finite-time {\it singularity} \cite{Nojiri:2005sx,Odintsov:2018uaw} where the Hubble function diverges, while the scale factors are always finite and nonvanishing. Such singular points will be carefully investigated in the next section studying the geodesic completeness and scalar perturbations across them. Here we just note that, in general, quantum effects of particle creation \cite{Montani:2001fp, Dimopoulos:2018kgl,Ford:1986sy,Contreras:2018two} can produce additional effective terms in the Friedman equation, able to regularize singularities of the Hubble function. 
\\Finally, the scalar fields behaviour is shown in Fig.~\ref{fig: PhiPsi} where the field $\phi$ asymptotically reaches unity as $t\to\infty$, while the Immirzi field relaxes to a constant Immirzi parameter.

\subsection{Radiation and dust}
\begin{figure}[h]
\centering
\includegraphics[width=2.5in]{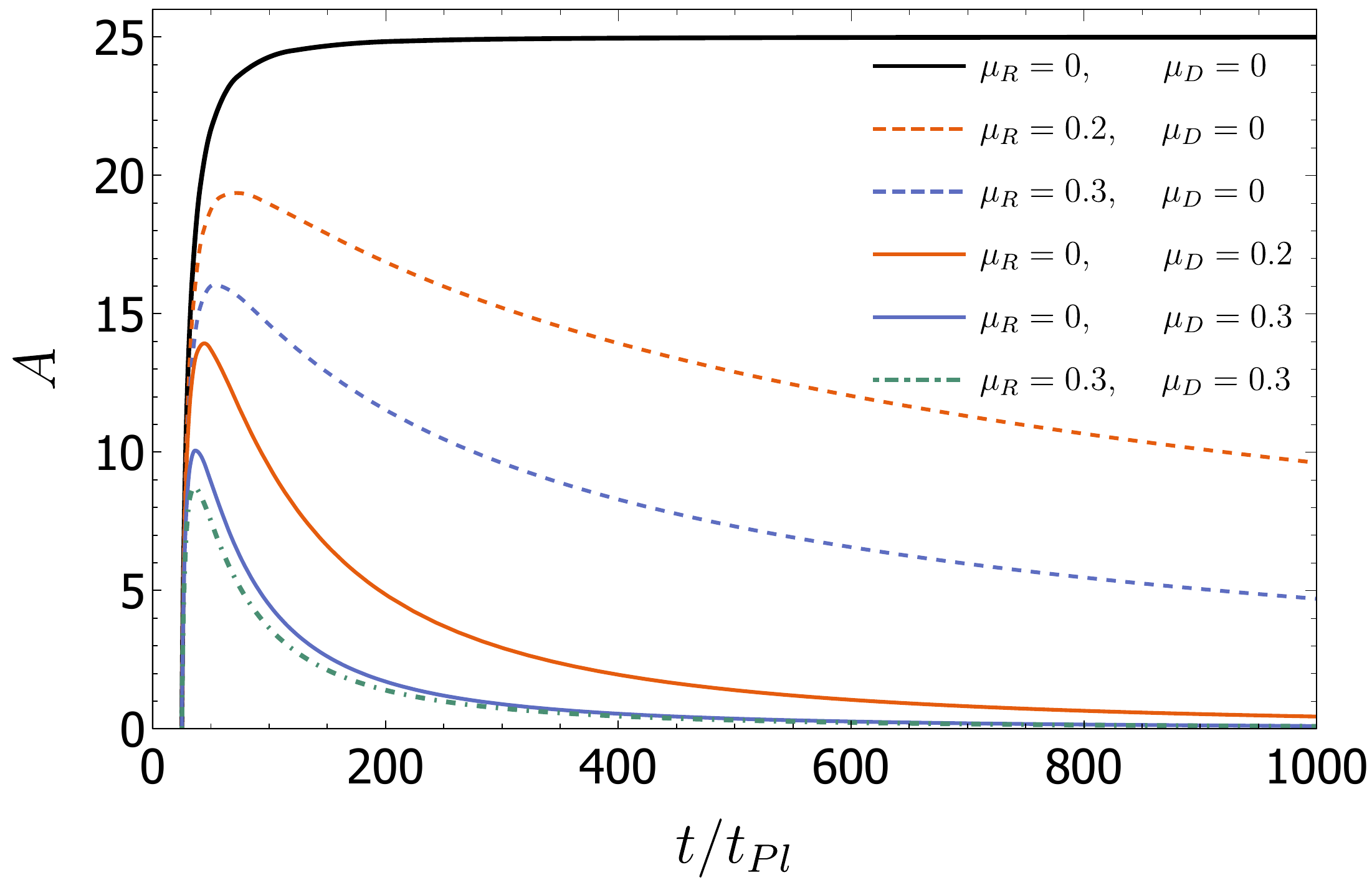}
\caption{Asymptotic behavior of the anisotropy degree $A$ as a function of $t/t_{Pl}$ after the finite time singularity for various values of $\mu_R$, $\mu_D$ and $\alpha=-5/3$, $\mu_I=\sqrt{3}$, $\mu_A=0.2 \mu_I$.}
\label{fig: Anisotropy}
\end{figure}

The above analysis can be extended to the non vacuum case, including the energy density of radiation and dust. A first difference concerns the late time region, since both radiation and dust are able to provide an isotropization mechanism for the universe, as is apparent studying the anisotropy degree
\begin{equation}
    A(t) = \frac{\left( H_A^2 + H_B^2 + H_C^2 \right)}{3 H^2} -1,
\end{equation}
whose behavior is shown in Fig.~\ref{fig: Anisotropy}.
Regarding the universe in its early phase, instead, two different scenarios may occur, depending on the value of the parameter $\alpha$. If $\bar{\alpha}<\alpha<0$, where $\bar{\alpha} = -2 \mu_I^2 / \mu_D^4$, the solutions are qualitatively equivalent to the vacuum case. If $\bar{\alpha}>\alpha$, instead, we still obtain a bouncing behavior for the volume (See Fig.~\ref{fig: norip}), which however is now devoid of singularities. However, now the scale factors either diverge or vanish at some critical time $t_c$.

\def\figsubcap#1{\par\noindent\centering\footnotesize(#1)}
\begin{figure}[h]%
\begin{center}
  \parbox{2.1in}{\includegraphics[width=2in]{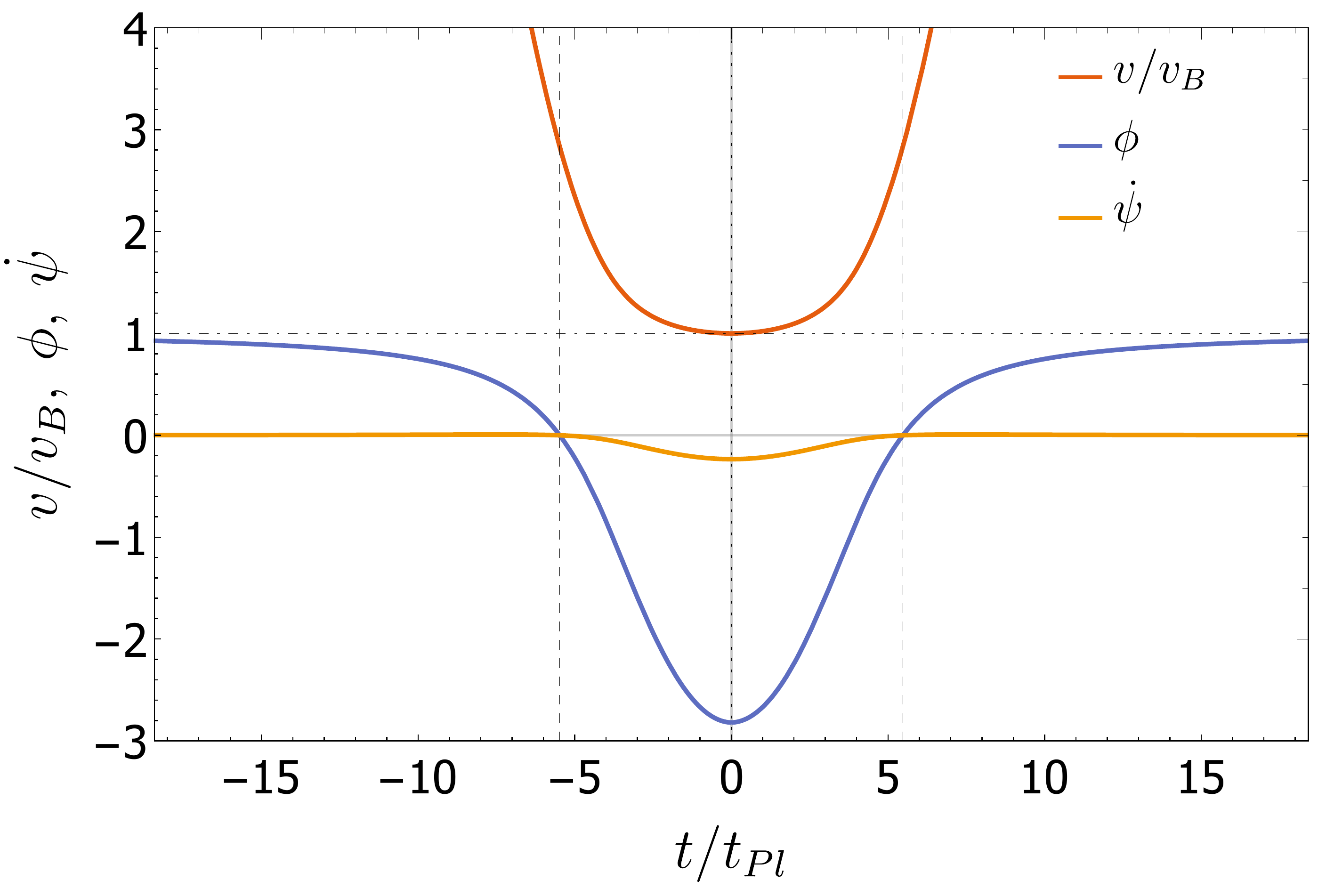}\figsubcap{a}}
  \hspace*{4pt}
  \parbox{2.1in}{\includegraphics[width=2in]{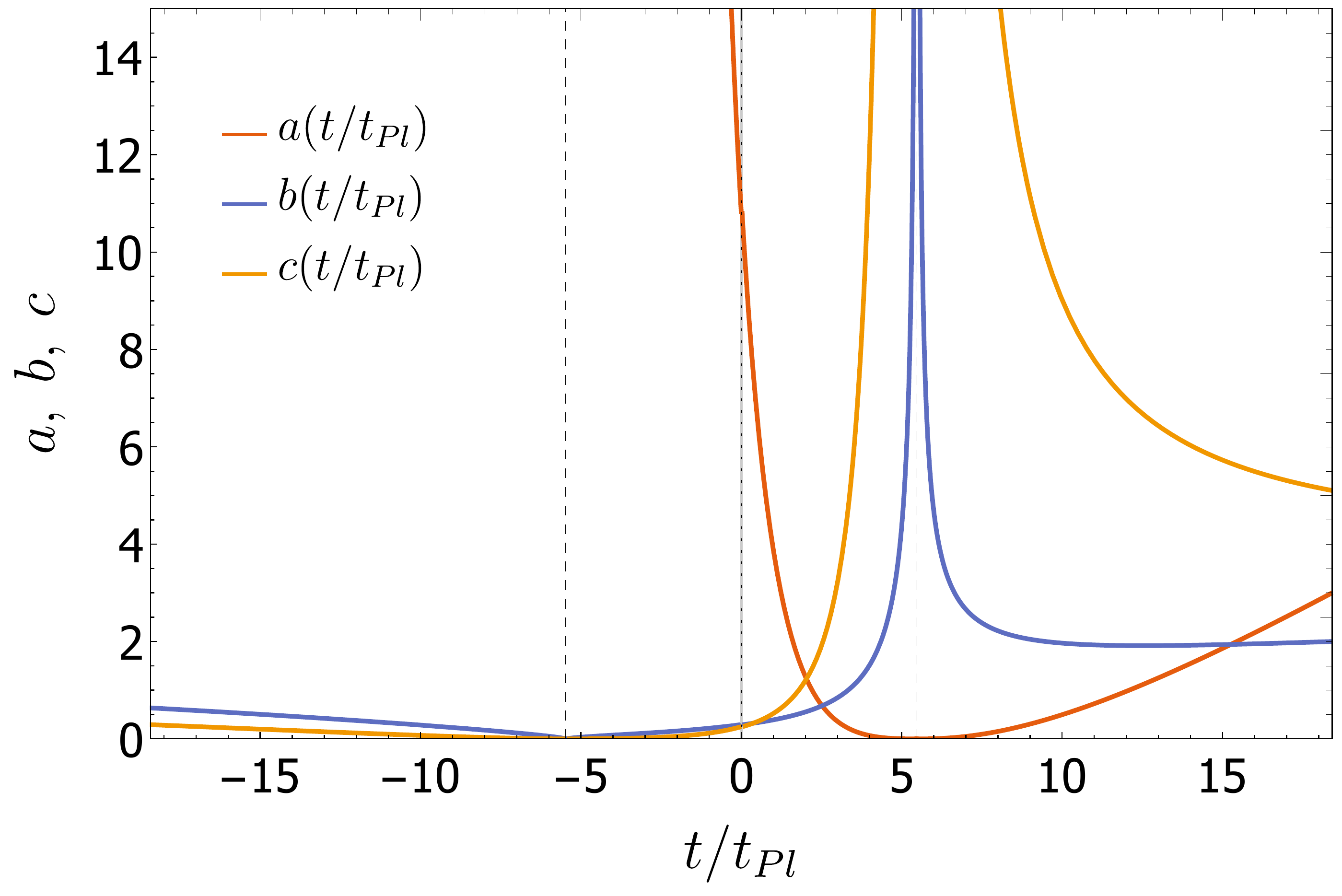}\figsubcap{b}}
  \caption{Numerical solutions as a function of $t/t_{Pl}$ for $\mu_I=0.057$, $\mu_A=2.4$, $\mu_D=0.365$, $\mu_R=1.56$ and $\alpha=-8.42<\bar{\alpha}$. The dashed lines represent where the scalaron vanishes. (a) Volume normalised to $v_B$, $\phi$ and derivative of the Immirzi field. (b) Scale factors.}%
  \label{fig: norip}
\end{center}
\end{figure}

\section{Physical implications of curvature divergences}\label{geodesiccompleteness}
As described in the previous section there are two classes of solutions, characterized by a singularity in the Hubble function and non vanishing and finite scale factors (Fig.~\ref{fig: PlotV}) or with regular Hubble function but vanishing or divergent scale factors (Fig.~\ref{fig: norip}). In this section we will analyse such singularities studying the behavior of null geodesics and of scalar perturbations near the critical time $t_c$, at which the singularity is located.

Regarding null geodesics with tangent vector $u^\alpha=dx^\alpha/ds$, it can be shown that they admit a first integral of the form \cite{Singh:2011gp,Nomura:2021lzz}
\begin{equation}
x'= \frac{k_a}{a^2}, \qquad y'= \frac{k_b}{b^2}, \qquad z'= \frac{k_c}{c^2}, \qquad t'= \left(\frac{k^2_a}{a^2}+\frac{k^2_b}{b^2} +\frac{k^2_c}{c^2}\right)^{1/2} +C_0,\label{geodesic derivative}
\end{equation}
where prime denotes derivative with respect to the affine parameter $s$ and ${k_a,k_b,k_c}$ and $C_0$ are integration constants. It follows that if $a(t)$, $b(t)$, and $c(t)$ are continuous and non-vanishing, as in Fig.~\ref{fig: PlotV}, the tangent vector to the geodesics will be unique and well defined. Therefore, such cases are geodesically complete, a result that holds both in the anisotropic and in the isotropic case \cite{Jimenez:2016sgs}.

\begin{figure}[h]
\centering
\includegraphics[width=2.5in]{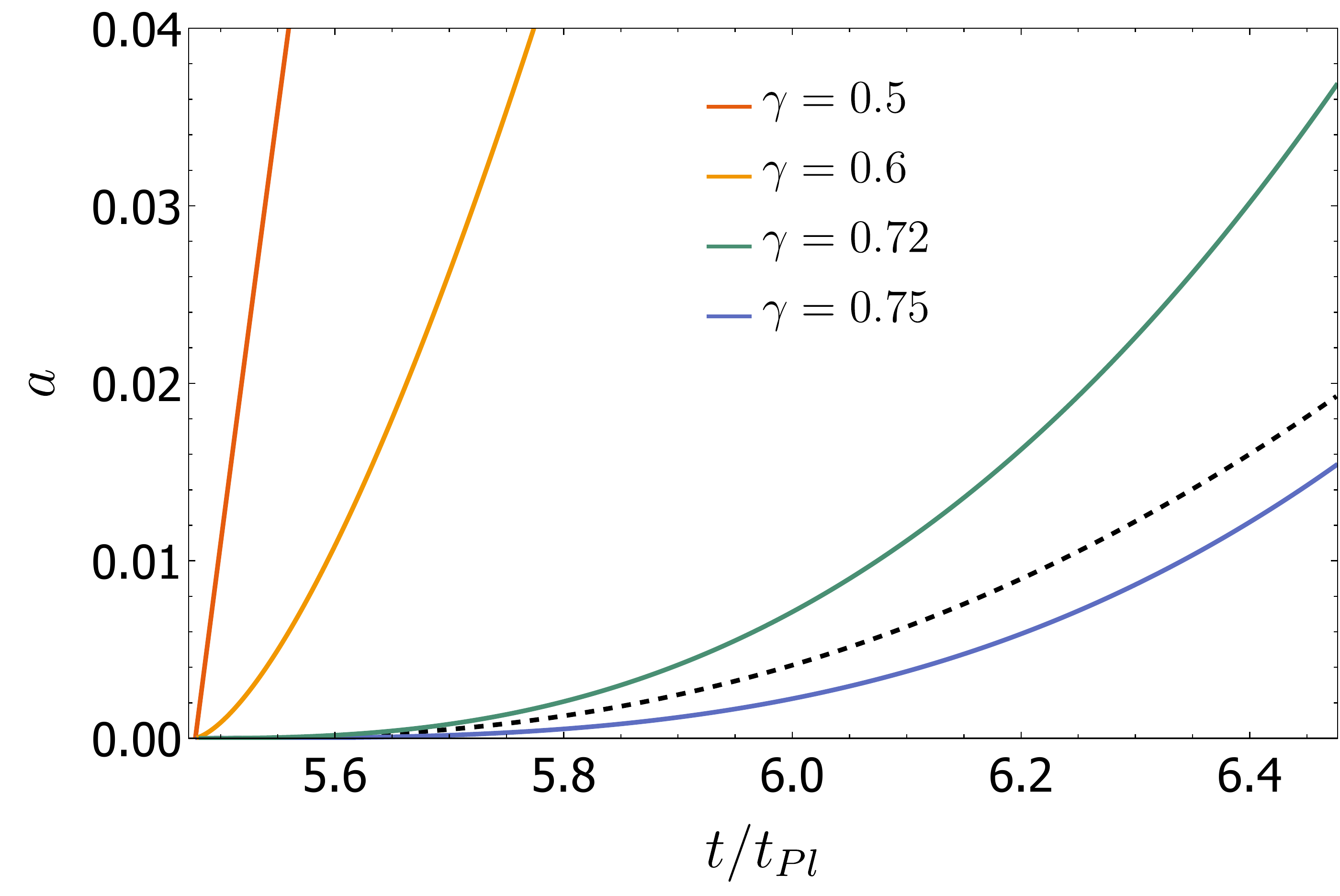}
\caption{Outcomes of null geodesics test for $\alpha<\bar{\alpha}$. Scale factor $a(s(t))$ for different values of $\gamma$. The dashed-black line represent the numerical solution $a(t)$ reported in Fig.~\ref{fig: norip}.}
\label{fig: geodesics}
\end{figure}

In the other class of solutions (Fig.~\ref{fig: norip}), instead, we see that the volume remains finite despite the vanishing/divergence of some scale factors. The divergence of individual scale factors does not affect the geodesics, but the vanishing of some of them may spoil the continuity and lead to the impossibility of a unique extension across $t_c$. Indeed, let us consider the case in which one scale factor vanishes at some affine parameter $s_c$. In particular, suppose that
\begin{equation}\label{geodesic scale factor}
    a(s)=a_0(s-s_c)^{\gamma},
\end{equation}
with $\gamma>0$. Then, integrating the relevant equations yields
\begin{align}
    x(s) &= x_c + \frac{k_a(s-s_c)^{1-2\gamma}}{a_0^2(1-2\gamma)},\\
     t(s) &= t_c + C_0(s-s_c) + \frac{k_a(s-s_c)^{1-\gamma}}{a_0(1-\gamma)},\label{geodesic time}
\end{align}
which are smooth if $0<\gamma<1/2$ and $0<\gamma<1$, respectively. Note that if $1/2<\gamma<1$, then $x(s)\xrightarrow[]{s\to s_c}\pm\infty$, which would imply reaching infinity in finite coordinate time. Conversely, if $0<\gamma<1/2$, then the geodesic path will span the range $\{t,x\}\in\; (-\infty,\infty )$, and the geodesics would be complete, despite the vanishing of some scale factors. However, this result is dependent on how rapidly the zero is reached, i.e. we have to determine the value of $\gamma$ relative to the solutions in question. The results are shown in Fig.\ref{fig: geodesics} and prove that the solution approaches zero too rapidly, corresponding to a value of $\gamma$ larger than $1/2$. We are thus forced to conclude that the example shown in Fig.~\ref{fig: norip} does represent a geodesically incomplete  space-time.

We now turn the attention to the behavior of scalar field perturbations. For a scalar mode given by $\sigma_{\vec{k}}(t,\vec{x})=\Theta(t) e^{i \vec{k}\cdot\vec{x}}$, one finds\cite{PhysRevD.103.124031} an equation of the form
\begin{equation}\label{eq:A-equation}
\ddot{\Theta}+h(v) \frac{\dot v}{v} \dot \Theta+\left(\frac{k_x^2}{a^2}+\frac{k_y^2}{b^2}+\frac{k_z^2}{c^2}\right)\Theta=0 \ , 
\end{equation}
where $h(v)$ represents some regular function of the volume $v$ and $\vec{k}=(k_x,k_y,k_z)$ represents a set of constants. From this expression, we see that scalar modes are sensible to the presence of the individual scale factors $a,b,$ and $c$, and of the Hubble function $H=\dot{v}/3v$.
If the scalar factors do not vanish anywhere, then any potential problems should come only from the term involving the Hubble function, which diverges in $t_c$.
In particular, in vacuum one finds\cite{PhysRevD.103.124031} the following  approximate solution near the singularity
\begin{equation}
\Theta(t)\approx \Theta_c+\frac{\dot \Theta_c}{2\tilde{h}_c^2}e^{\mp2\tilde{h}_c|t-t_c|^{1/2}}\left(1\pm2\tilde{h}_c|t-t_c|^{1/2}\right) \ ,
\end{equation}
with $\Theta_c, \dot \Theta_c, \tilde{h}_c$ constants, from which we see that scalar field perturbations remain bounded around $t_c$ despite the divergence in the Hubble function. A similar result holds in presence of dust and radiation, since
\begin{align}
    \Theta \approx & \Theta_c- \frac{\dot{\Theta}_c}{32 \tilde{h}_c^4} e^{\mp4\tilde{h}_c|t-t_c|^{1/4}} \left(  3\pm12\tilde{h}_c|t-t_c|^{1/4} \right.\nonumber\\
    &\left.+24\tilde{h}_c^2 |t-t_c|^{1/2} \pm 32\tilde{h}_c^3 |t-t_c|^{3/4} \right),
\end{align}
which is easy to see to be again bounded.
On the other hand, for a regular Hubble function but vanishing scale factors, equation \eqref{eq:A-equation} describes a harmonic oscillator with a time dependent frequency, 
\begin{equation}
\ddot{\Theta}(t)+\frac{k_x^2}{a^2(t)}\Theta(t)\approx0,
\end{equation}
which diverges as $a(t)\to 0$. Therefore, for the values of $a(t)$ obtained numerically in the previous section neither geodesics nor scalar perturbations are well behaved.

\section{Conclusion}\label{conclusions}
We proposed a generalization of the Nieh-Yan term to metric-affine gravity, by including an additional term featuring nonmetricity and inserting two parameters ($\lambda_1$, $\lambda_2$), which allow to recover the projective invariance and the topological character, otherwise lost in presence of nonmetricity. In particular, projective invariance can be independently obtained by setting $\lambda_1=\lambda_2=\lambda$, whereas topologicity is only guaranteed for $\lambda=1$.
\\As an explicit example, we considered a model with Lagrangian $F(R,NY_{gen})$, which re-expressed in the Jordan frame features two scalar fields. We identified these scalar degrees as the $f(R)$-like scalaron $\phi$ and the Immirzi field $\beta$ and showed that the latter acquires dynamical character and a potential term in a more natural way than in previous treatments, where these features are introduced by hand in the action. Depending on the values of $\lambda_1$ and $\lambda_2$, we found two different scalar tensor theories.
Models with $\lambda_1\neq\lambda_2$, are non-dynamical, and the scalar fields are frozen to constant values in vacuum.
In the projective invariant case ($\lambda_1=\lambda_2$), instead, the theory is endowed with one additional dynamical degree of freedom, the Immirzi field, while the field $\phi$ is algebraically related to the latter via a modified structural equation. 
\\ Then, we considered in more detail the dynamical models, looking for cosmological solutions in Bianchi I spacetimes. We found numerical solutions in which the big bang singularity is replaced by a big bounce scenario, in which the universe volume undergoes a contraction up to a minimum value and then bounces back, re-expanding in another branch.  The scalaron and the Immirzi field reach a maximum during the bounce and relax to constant values at later times, where the standard LQG picture, with $\phi=1$ and a constant Immirzi parameter $\beta=\beta_0$, is recovered. Moreover, the inclusion of dust and radiation turns out to provide an isotropization mechanism at late times, a feature that is absent in the vacuum case.

Such solutions are characterized by future finite time singularities after the bounce, either in the Hubble function or in the individual scale factors.
In the former case, we showed that null geodesics are still well behaved and scalar perturbations bounded, which allows us to conclude that the solution is physically acceptable. In the latter case, instead, the study of null geodesics shows that they cannot be extended across the singular point, where also scalar perturbations grow in time, leading us to regard such solutions as unphysical.

\bibliographystyle{ws-procs961x669}
\bibliography{references}

\end{document}